\documentclass[onecolumn,showpacs,preprintnumbers,amsmath,amssymb]{revtex4}

\usepackage{graphicx}
\usepackage{dcolumn}
\usepackage{bm}

\begin{document}

\preprint{APS/123-QED}

\title{Special Relativistic Liouville Equation Completed} 

\author{Jose A Magpantay}
\affiliation{Quezon City, Philippines}
\email{jose.magpantay11@gmail.com}

\date{\today}

\begin{abstract}
In two previous papers, the author raised the possibility of a special relativistic Liouville equation. The conclusion then was yes, such an equation is possible in 8N phase space if a Lorentz-invariant, Universal (LiU) time can be defined for all the degrees of freedom. Without this LiU time, the existence of a special relativistic Liouville equation is empty and may just be a waste of time. In this paper, I propose and argue that the LiU time follows from entropy, which should not be surprising given the second law of thermodynamics and the fact that regardless of how temperature and heat transforms under Lorentz transformation the entropy is Lorentz invariant. Thus, it is natural to define LiU time from entropy. This now completes the existence of a special relativistic Liouville equation, which will determine the Gibbs distribution, the starting point of classical statistical mechanical description of physical systems. To illustrate the formalism, the partition function for the relativistic ideal gas is derived. The result is much simpler than the Juttner gas result.     
\end{abstract}

\pacs{Valid PACS appear here}
\maketitle

1. A special relativistic classical statistical mechanics is accepted in terms of a Boltzmann equation since it involves a single particle distribution, and the terms in the equation, including the collision term, can be made relativistic \cite{de Groot}. But for the full N particle Gibbs  distribution given by the Liouville equation, this is problematic for two reasons. First, the old theorem of (1) Currie, Jordan and Sudarshan \cite{Currie} ; (2) Cannon and Jordan \cite{Cannon} and (3) Leutwyler \cite{Leutwyler} states there is no relativistic Hamiltonian for interacting particles because no Hamiltonian, the time translation generator, will satisfy the Poincare algebra. Without a Hamiltonian, there is no Liouville equation. Second, which time do we use in the Liouville equation since all the particles will have their own relativistic time. 

In a previous paper \cite{Magpantay1}, I showed how a Hamiltonian for a non-relativistic many particle system with two body-interaction can be made relativistic. The Hamiltonian involves a time-delay in two body interaction, which was also shown to cause the Hamiltonian not satisfying the equal time commutator inherent in the Poincare algebra. Regardless, the resulting Hamiltonian being relativistic is a viable Hamiltonian, albeit with many relativistic times for each of the particles. Now we also have the problem of which time do we use in a Liouville equation for the Gibbs distribution of N particles, which must make use of a single relativistic time. In another paper \cite{Magpantay2}, I argued that the best way to have a relativistic Liouville equation is to formulate statistical mechanics in 8N phase space by introducing a mathematical time t for all the particles. For Lorentz invariance of the Hamiltonian, we find that the mathematical time must be a Lorentz scalar. But the Hamiltonian has a new symmetry under the mathematical time, it is invariant under reparametrization of time t. Applying the Dirac constraint formalism \cite{Dirac}, we find that the reparametrization symmetry leads to a constraint, the Hamiltonian is zero, which is similar to general relativity's Hamiltonian is zero due to diffeomorphism symmetry. Thus, we seem to be led to a trivial Hamiltonian, and no Liouville dynamics for the Gibbs distribution.

The third paper \cite{Magpantay3} on this problem fortunately showed that there is a way to break the reparametrization symmetry, which will make the velocities solvable in terms of the momenta.  It also leads to a Hamiltonian simply factored by a constant, which measures the reparametrization symmetry breaking, in a way that the Liouville equation can absorb the constant in its partial derivative with respect to the Lorentz scalar time t of the Gibbs distribution. Thus, the Hamiltonian exists, and the mathematical time dependence is simply scaled by the symmetry breaking constant. There is a special relativistic Liouville equation for the Gibbs distribution. But the whole thing depends on the mathematical time introduced. And this time must be the same for all N particles, thus it must be universal and it must be a scalar, thus it must be Lorentz invariant. Can such a Lorentz-invariant Universal (LiU) time exist and can it be used to "time" the N particle statistical dynamics?

2. When I was working on this problem before, the answer should have been obvious, the time needed is defined by entropy because of the second law of thermodynamics and by the fact that all groups that worked on special relativistic thermodynamics agreed on one thing, that regardless of how temperature and heat transform under Lorentz transformation, entropy is Lorentz invariant. But I missed it then. In this section, I present the arguments why entropy is Lorentz invariant. 

The first group is led by Planck and followed by Einstein (for a while) and by Juttner, which says that 
\begin{subequations}\label{1}
\begin{gather}
T = (1 - \frac{u^{2}}{c^{2}})^{\frac{1}{2}} T_{0}, \label{first} \\
Q =  (1 - \frac{u^{2}}{c^{2}})^{\frac{1}{2}} Q_{0},
\end{gather}
\end{subequations}
where $ Q_{0} $ and $ T_{0} $ are the heat and temperature in the rest or proper frame of the thermodynamic system and  Q and T are in the frame where the thermodynamic system moves along the $ x $ axis with velocity u. From equation (1), it follows from the second law of thermodynamics that entropy is Lorentz invariant. 

The derivation by Planck is a bit involved as argued in \cite{Balescu} and a simpler derivation, rather straightforward, is given in \cite{Tolman}. Tolman's derivation follows from the relativistic transformation of mechanics and the first law of thermodynamics,
\begin{equation}\label{2}
dE = dQ - dW,
\end{equation}
 i.e.,  change in internal energy of a thermodynamic system is equal to heat transferred minus the work done. However, Tolman argued the statistical interpretation of entropy, that it is the probability of finding it in a given state should not depend on the Lorentz frame, i.e., it is Lorentz invariant. Using the second law of thermodynamics,
\begin{equation}\label{3}
dS \geq \dfrac{dQ}{T},
\end{equation}
derived the temperature transformation given by equation (1a).   

Equation (1a) shows that the temperature of a moving body is colder. This does not seem to follow from the relativistic transformation of velocities and equipartition theorem, thus implying higher temperature per particle. A number of physicists, including Ott, Kibble and Moller proposed instead the transformations, see for example the review in \cite{Requardt}  
\begin{subequations}\label{4}
\begin{gather}
T = (1 - \frac{u^{2}}{c^{2}})^{\frac{-1}{2}} T_{0}, \label{first}\\
Q =  (1 - \frac{u^{2}}{c^{2}})^{\frac{-1}{2}} Q_{0}. 
\end{gather}
\end{subequations}

There is a third transformation \cite{Landsberg}, which states that both temperature and heat are scalars, and are thus invariant. Again, entropy is invariant under this relativistic thermodynamics.

These transformations show relativistic thermodynamics is rather confusing because the variables are non-local making its definition and measurement rather ambiguous \cite{Dunkel}. It is believed that the problem is nowhere near resolution \cite{Farias}, and \cite{Wang} claims that Einstein took all three positions in certain stages of his life. The only point I want to emphasize is that in spite of all these confusion, thermodynamics entropy is a Lorentz invariant quantity and was shown directly to be Lorentz invariant in a covariant statistical and thermodynamics formulation \cite{deParga}. And given the second law of thermodynamics, i.e., the entropy of an isolated system always increases, then entropy is argued by some \cite{Grandy} as time. More recently, a geometric proof that time is entropy is given in \cite{Quevedo}. However, this proof is only rigorously shown for an ideal gas. Regardless, it can be argued that all these hints that defining a Lorentz-invariant Universal (LiU) time needed to implement a special relativistic Liouville equation can be done.  

3. To use entropy as a basis for LiU time t means determining $ t = t(S) $, where S is entropy, and that is a big problem. The system must be closed, not subject to, outside influence, so that entropy keeps increasing for a reasonable length of time and expect t to keep on increasing with it. The only closed system is the entire universe (multiverses are not considered here) and the accepted cosmology begins with a primordial atom with very low entropy (can be zero if only a singlet state), the universe evolved into its current state after 13.8 B years with very high entropy that we have no idea how to compute from current theories. Thus, we only get a very coarse time line of the universe's evolution and a truely universal LiU $ t(S) $ is but a dream. 

If we now consider a finite system, the evolution of structures like in living systems, can lead to the lowering of entropy of the system for some time. But there is no violation of the second law of thermodynamics, the interaction with the environment, guarantees it. But does it mean that a LiU time for such a system runs backwards during entropy decrease? Apparently not, the living system still 'ages'. Maybe we just have to forego the idea of entropy as a basis for LiU time for such finite systems. 

In the laboratory, an experiment where a system, say a volume of gas molecules, can be isolated and observed to evolve. Calorimetry and thermal measurements can be made without interfering with the system's processes. Is it possible to define a $ t(S) $ in the laboratory? Of course this can only be done while the system moves to equilibrium, when entropy achieves its maximum and stays there, and from then on LiU time stops. The simplest dependence may be given by      
\begin{equation}\label{5}
t = \alpha S,
\end{equation}
where t is in seconds and S must be in calorie or joule per temperature measured in kelvin K. The proportionality constant $ \alpha $ must be in seconds times unit of  $ [ entropy ]^{-1}  $. Equation (5) guarantees a LiU time. But how do we determine $ \alpha $ ? This can be done by doing a measurement in two events at times $ t_{1} $ and $ t_{2} $. Not having a LiU time clock yet, we have to make do with a precise atomic clock that is tied to measurements of the gas' thermal measurements. It is reasonable to take a temperature difference of 1 K between the two measurement events that may differ in time by an order of seconds. Since a typical particle in a thermodynamic system at temperature T has energy equal to $ kT $, where k is the Boltzmann constant (unit of $ joules/kelvin $), then a unit of entropy of a thermodynamic system must be equal to $ k $ $(1 Kelvin) $ $ N $ in joules, where N is the total number of particles, typically Avogadro's number, giving 
\begin{equation}\label{6}
\alpha = \frac{1}{1.38} \dfrac{seconds}{joule{\frac{1}{K}}}, 
\end{equation}
which says time and entropy are roughly of order 1. Thus, in a laboratory setting such as this one, entropy can measure time at the same order of magnitude.

Next, LiU time has a reparametrization symmetry and must be accounted for. At the level of the entire universe, it is to be emphasized that LiU time is not the same as the time component of the space-time, which is invariant under general coordinate transformation. As discussed at the beginning of this section, our current physics does not allow us to compute the entropy and we can only give a very coarse grained evolution of the universe. But there are many degrees of freedom (quarks, leptons and vector fields) with various symmetries (gauge, global symmetries that lead to conservation laws, space-time symmetries, discrete symmetries that lead to multiplicative conserved quantities) that surely would lead to a reparametrization symmetry of the entropy. Thus, a reparametrization symmetry of the universe's  LiU time. 

How can such reparametrization symmetry appear in entropy measurements in the laboratory? The peculiarity of the entropy measurement depends on the thermodynamic system. So for simplicity, consider a single type of gas, contained in a volume V, pressure p, temperature T. From \cite{Reif}, the thermodynamic potentials E(S,V) (free energy), H(S,p) (enthalpy), F(T,V) (Helmholz free energy) and G(T,p) (Gibbs free energy) are related to each other, starting from the free energy given by equation (2) with $ dW = pdV $ by
\begin{subequations}\label{7}
\begin{gather}
H = E + pV, \label{first} \\
F = E - TS, \label{second} \\
G = E - TS +pV. 
\end{gather}
\end{subequations}
From equations (7) and (2), the relevant differential equation for entropy are
\begin{subequations}\label{8}
\begin{gather}
dF = -S dT - p dV,\label{first}\\
dG = -S dT + V dP,
\end{gather}
\end{subequations}
from which we read
\begin{subequations}\label{9}
\begin{gather}
S = - (\dfrac{\partial F}{\partial T})_{V = const}, \label{first}\\
S = - (\dfrac{\partial G}{\partial T})_{p = const}.
\end{gather}
\end{subequations}. 

Equations (9a,b) suggests how reparametrization symmetry in a laboratory entropy measurements can arise - the measurement of entropy via F or via G give two reparametrizations already. And there may be more reparametrizations depending on the specific experimental set-ups. There will also be additional variations if we add other thermodynamic quantities, like other  types of gases, which will add moles and number of molecules for each type of gas in the variables for the thermodynamic potentials. In short, I have argued how reparametrization symmetry can arise in entropy measurements in a laboratory.    

Finally, can entropy measurements give a reasonably fast time resolution, maybe in seconds or fraction of a second? If the time resolution determined by entropy measurements is in minutes to hours, it will not be of much use as a clock. My knowledge of experimental techniques and tools is almost non-existent for me to give a sensible guess, such as the one I made in this section.  

4. To illustrate how the ideas in SR Liouville may be used, I will discuss an example, the Juttner gas - particles are relativistic but do not have interaction. A relativistic particle will have energy given by 
\begin{equation}\label{10}
E = \sqrt{c^{2}\vec{p}\cdot\vec{p} + m^{2} c^{4} }. 
\end{equation}
For N non-interacting particles,  the energy is just the sum of equation (10) for all particles, which gives the free relativistic particles partition function in 6N phase space as  
\begin{equation}\label{11}
\begin{split}
Z_{J}& = \frac{1}{N!}(\frac{1}{h^{3}})^{N} \int \prod_{a=1}^{N} d\vec{p}^{a} d\vec{x}^{a} \exp {(-\beta \sum_{a=1}^{N} c(\vec{p^{a}}^{2} + m^{2}c^{2})^{\frac{1}{2}})}\\
			&  = V^N \frac{1}{N!}(\frac{1}{h^{3}})^{N} \left[ 4\pi \frac{m^{2}c}{\beta} K_{2}(mc^{2}\beta) \right]^{N},
\end{split}
\end{equation}
where $ \beta = \frac{1}{kT} $, $ V $ is the volume of the container, $ h $ is the unit of volume in phase space and $ K_{2} $ is a modified Bessel function of the second kind of order 2 and the subscript J in Z is to emphasize it is the result attributed to Juttner \cite{Chacon-Acosta}.  

But in writing down equation (11), use is made of the Gibbs distribution function, which satisfies
\begin{equation}\label{12}
\dfrac{\partial f}{\partial t} - \left\lbrace H, f \right\rbrace  = 0.
\end{equation}
The questions in equation (12) are first, which time is being referred to in the time derivative and the equal time Poisson bracket with H since there are N particles, each with their own relativistic time. Second, thermal equilibrium necessarily implies $ \left\lbrace  H, f \right\rbrace  = 0 $, thus $ f \sim H $. But there are N particles with momentum for each particle defined using its own relativistic time. So, the problem of time is still there. But this is usually finessed by assuming there is one relativistic time for all the particles because they are non-interacting anyway. This subtle move results in $ f = H $ , with  
\begin{equation}\label{13}
H= \sum_{a} \sqrt{ \vec{p_{a}}\cdot\vec{p_{a}} c^{2} + m^{2} c^{4} },
\end{equation}
and this results in the partition function given by equation (11). 

In this section, I will try to derive the partition function for non-interacting relativistic particles, by taking into account the results of the proposal for a LiU time and the proposed Liouville equation in 8N phase space \cite{Magpantay3}. Here there is no problem with which time is being used because all particles are making use of the LiU time. I will give the corrected relevant formulas from that reference because there are algebraic mistakes in the derivation of the 8N Hamiltonian in that paper.  
\begin{subequations}\label{14}
\begin{gather}
S_{r,\gamma} = \int dt L_{r,\gamma}, \label{first}\\
\begin{split}
L_{r,\gamma}& = \sum_{a} \Big\lbrace (-mc) \left( \eta_{\mu \nu} \dfrac{dx_{a \mu}}{dt}  \dfrac{dx_{a \nu}}{dt} \right)^{\frac{1}{2}} + \gamma \left( \eta_{\mu \nu} \dfrac{dx_{a \mu}}{dt}  \dfrac{dx_{a \nu}}{dt} \right)\Big\rbrace.
\end{split}
\end{gather}
\end{subequations} 
Note that the first term of equation (14b) gives the reparametrization symmetric term while the $ \gamma $ term breaks the reparametrization symmetry, which allows for the computation of the velocities in terms of the momenta as given by
\begin{subequations}\label{15}
\begin{gather}
\dfrac{dx_{a0}}{dt} = \frac{1}{2\gamma} \left[ 1 - mc\left( \pi_{a}^{2} - \vec{p}_{a} \cdot \vec{p}_{a} \right)^{\frac{-1}{2}} \right] \pi_{a},\label{first}\\
\dfrac{d\vec{x}_{a}}{dt} = - \frac{1}{2\gamma} \left[ 1 -  mc\left( \pi_{a}^{2} - \vec{p}_{a} \cdot \vec{p}_{a} \right)^{\frac{-1}{2}} \right] \vec{p}_{a}.
\end{gather}
\end{subequations} 
Note that $ \gamma $ is not dimensionless, equations (14b) and (15) show the dimension of $ \gamma $, usually represented by $ [ \gamma ] $ is mass. 

The free particles relativistic Hamiltonian in 8N phase space, taking into account the metric $ (+,---) $ is
\begin{equation}\label{16}
\begin{split}
H_{fr}& = \sum_{a} \left( \pi_{a}\dot{x_{0a}} - \vec{p_{a}} \cdot \dot{\vec{x_{a}}} \right) - L_{r,\gamma} \\
        & = \frac{1}{\gamma} \sum_{a} F(\pi_{a}, \vec{p_{a}}), \\
\end{split}
\end{equation}
and $ F(\pi_{a}, \vec{p_{a}}) $ is given below as
\begin{equation}\label{17}
F(\pi_{a}, \vec{p_{a}}) = \frac{1}{4} \Big\lbrace \left[ \left( \pi_{a}^{2} - \vec{p_{a}}
\cdot \vec{p_{a}} \right)^{\frac{1}{2}} - mc \right]^{2} + 4 \vec{p_{a}} \cdot \vec{p_{a}}
\left[ 1 - mc \left( \pi_{a}^{2} - \vec{p_{a}} \cdot \vec{p_{a}} \right)^{\frac{-1}{2}} \right] \Big\rbrace.
\end{equation}
Since $ \gamma $ has a dimension of mass, $ H_{fr} $ correctly has dimension of energy. 

I begin from the 8N phase space partition function given by 
\begin{equation}\label{18}
Z_{fr} = \frac{1}{N!}(\frac{1}{h_{0}^{4}})^{N}\prod_{a} \int d\vec{x_{a}} d\vec{p_{a}} dx_{0a} d\pi_{a} \exp { (-\beta H_{fr}) },
\end{equation}
which makes $ \beta H_{fr} $ dimensionless since $ \beta = \frac{1}{kT} $. Also, $ h_{0} $ is the smallest volume in phase space occupied by the particles. Unfortunately, the integral over $ \pi_{a} $ cannot be found in books of Table of Integrals. But a suitable solution is suggested by the fact that $ \gamma \rightarrow 0 $ in $ H_{fr} $, which is also desired to restore the reparametrization symmetry in $ L_{r,\gamma} $ in equation (14). Also, the Liouville equation for the Gibbs distribution see \cite{Magpantay3} says as $ \gamma \rightarrow 0 $, the Gibbs distribution attains equilibrium, which says $ \left\lbrace H_{fr},f \right\rbrace  = 0 $, which makes $ f = H_{fr} $ giving equation (18). Because of this, I can make use of
\begin{equation}\label{19}
\delta_{\alpha}(x) = \frac{1}{\sqrt{\pi \alpha}} \exp {(-\frac{x^{2}}{\alpha})},
\end{equation}
with $ \alpha \rightarrow 0 $. This is also used extensively in gauge-fixing in gauge theories. I write $ F(\pi_{a},\vec{p_{a}}) = ( \sqrt{F(\pi_{a},\vec{p_{a}})} )^{2} $, then use can be made of 
\begin{equation}\label{20}
lim_{\gamma \rightarrow 0} \frac{1}{\sqrt{\pi \gamma}} \exp {[-\frac{1}{\gamma} \beta F(\pi_{a},\vec{p_{a}})]} = \delta(\sqrt{\beta F(\pi_{a},\vec{p_{a}})}).
\end{equation}
But 
\begin{equation}\label{21}
\delta(\sqrt{\beta F(\pi_{a},\vec{p_{a}})}) = \dfrac{\delta(\pi_{a} - \tilde{\pi_{a}})}{\vert \beta^{\frac{1}{2}} \dfrac{\partial F^{\frac{1}{2}}}{\partial \pi_{a}} \vert_{{\pi_{a}} = \tilde{\pi}_{a} } },
\end{equation}
where $ \tilde{\pi}_{a} $ is a solution of $ F(\tilde{\pi}_{a},\vec{p_{a}}) = 0 $. Equation (16b) gives
\begin{equation}\label{22}
\begin{split}
\tilde{\pi}_{a}& = \sqrt{\vec{p}_{a}\cdot\vec{p}_{a} + m^{2} c^{2} } \\ 
                    & = \frac{E_{a}}{c},
\end{split}
\end{equation}
essentially giving the relativistic energy. The denominator of equation (21) is given by
\begin{equation}\label{23}
\beta^{\frac{1}{2}} \dfrac{\partial F^{\frac{1}{2}}}{\partial \pi_{a}} \vert_{\pi_{a} = \tilde{\pi}_{a}}  = \left[ 2\beta^{\frac{1}{2}} \dfrac{\vec{p_{a}}\cdot\vec{p_{a}}}{m^{2} c^{2}} \sqrt{\vec{p_{a}}\cdot\vec{p_{a}} + m^{2} c^{2}} \right] \left( F^{\frac{-1}{2}}(\tilde{\pi_{a}}, \vec{p_{a}} \right) .
\end{equation}
The use of $ F(\tilde{\pi_{a}}, \vec{p_{a}}) = 0 $ will be made later. 
Using equations (20) to (23), the partition function is now given by
\begin{equation}\label{24}
Z_{fr} = \frac{1}{N!}(\frac{1}{h^{4}})^{N} \prod_{a} d\vec{p_{a}} d\vec{x_{a}} dx_{0a} \left[ \sqrt{ \pi \gamma }  F^{\frac{1}{2}}(\tilde{\pi}_{a}, \vec{p}_{a}) \right] \left[ 2\beta^{\frac{1}{2}} \dfrac{\vec{p}_{a}\cdot\vec{p}_{a}}{m^{2} c^{2}} \sqrt{ \vec{p}_{a}\cdot\vec{p}_{a} + m^{2} c^{2} } \right]^{-1}.
\end{equation}
There are two terms that go to zero in above, these are $ \gamma $ and $ F(\tilde{\pi}_{a},\vec{p}_{a}) $. Let us deal with these terms that go to zero. From equation (16), the dimension of $ F(\pi_{a}, \vec{p}_{a}) $ is $ ( mass \dfrac{length}{time} )^{2} $ while the dimension of $ \gamma $ is mass. The integral over $ \vec{p}_{a} $ can be carried out. This is rather simple and involves, after change of variables $ p_{a} = (mc) tan\theta $, where $ \theta $ is in the range $ (0,\frac{\pi}{2}) $ resulting in 
\begin{equation}\label{25}
\int_{0}^{\frac{\pi}{2}} d\theta \sec \theta = (\ln 2 - \ln \cos \theta_{0}),
\end{equation}
where $ \theta_{0} \rightarrow \frac{\pi}{2} $, which will make the second term of equation (25) logarithmically divergent. But we can remove this undesirable divergent term by remembering the ignored terms in calculating $ Z_{fr} $, the $ \sqrt{\pi \gamma} \rightarrow 0 $ and $ F(\pi_{a}, \vec{p_{a}}) \rightarrow 0 $ in equation (25). So for each a, these two terms dominate the logarithmically divergent term $ - \ln \cos \theta_{0} $ as  $ \theta_{0} \rightarrow \frac{\pi}{2} $, because two power laws going to zero even with power $ \frac{1}{2} $ will dominate a divergent logarithmic term and thus the undesirable divergent term can be made to go away. This trick is a bit of cheating? No, because the $ \gamma \rightarrow 0 $ and $ F(\pi_{a}, \vec{p_{a}})\rightarrow 0 $ will now be accounted for in the finite term of equation (25), the $ \ln 2 $ term, in the remaining integrals in $ x_{0a} $ and $ \vec{x_{a}} $. Dimensionally, $ [ dx_{0a} ] $ and $ [ d\vec{x_{a}} ] $  are lengths. Taking into account the dimensions of $ \gamma $ and  $ F(\pi_{a}, \vec{p_{a}}) $, a new length differential can be defined as
\begin{equation}\label{26}
d\tilde{x_{0a}} = d\left[  \dfrac{\gamma^{\frac{1}{8}} F^{\frac{1}{8}}(\pi_{a}, \vec{p_{a}})}{\left( m^{\frac{3}{8}} c^{\frac{1}{4}} \right) } \right] x_{0a} .
\end{equation}
The powers in $ \frac{1}{8} $ arises because the zero approaching terms will be distributed in the four space-time terms. The m and c terms are introduced to cancel the dimensions coming from $ \gamma $ and $ F(\pi_{a}, \vec{p_{a}}) $. The effect is the new coordinates will have dimensions of length also. The m and c terms introduced must be cancelled by multiplying the $ Z_{fr} $ by $ m^{\frac{3}{2}} c $ for each particle. Since the mass of the molecule is typically of the order of $ 10^{-26} $ and c is of the order of $ 10^{8} $, the extra terms from m and c of equation (26) is of the order of $ 10^{8} $. The $ \gamma \rightarrow 0 $ and $ F(\pi_{a}, \vec{p_{a}})\rightarrow 0 $ makes the new coordinate differentials small, as differentials should be. 
Taking all these into account, the partition function is now given by 
\begin{equation}\label{27}
Z_{fr} = \frac{1}{N!} (V cT)^{N} (\frac{1}{h^{4}})^{N} \left( \dfrac{2\pi \ln 2  m^{\frac{7}{2}} c^{3}}{\beta^{\frac{1}{2}}} \right)^{N}.
\end{equation}
Note, this is a lot simpler than equation (11), the Juttner result. And this result was derived by making use of a LiU time, which makes the 8N Hamiltonian and the Liouville equation consistent special relativistically.  

5. To summarize, the Special Relativistic Liouville equation is completed now with the identification of entropy with time. This was facilitated by the fact that regardless of which version of relativistic thermodynamics is used, entropy is Lorentz invariant. Also, there have been philosophical arguments and at the moment limited mathematical proof that entropy is time. The reparametrization symmetry that arise due to the introduction of a LiU time is also argued to arise from the different thermodynamic potentials and by adding other thermodynamic intensive and extensive quantities. But the author's limitation in experimental techniques left many important questions on the possibility of entropy 'timers', the smallest unit of time that entropy can measure. Finally, the simple example a free relativistic gas was used to show that the Special Relativistic Liouville equation that makes use of a Lorentz-invariant Universal time can give results. The result is even much simpler compared to the Juttner gas, which has a drawback on what time is used. In principle, the use for interacting particles is also well-defined. But the evaluation of integrals may just be too complicated.

\begin{acknowledgements}
I would like to thank Felicia for correcting my Latex file. To Gravity, thank you for teaching One how to keep me company, which made working on this paper free of distractions.
\end{acknowledgements}


\begin{thebibliography}{20}
\bibitem{de Groot} 
de Groot S. R., W. A. van Leewen and Ch. G. van Weert, Relativistic Kinetic Theory, North Holland, Amsterdam, 1980.
\bibitem{Currie}
Currie, D. G., Jordan, T. F., and Sudarshan, E. C. G., Relativistic Invariance and Hamiltonian Theories of Interacting Particles, Rev. Mod. Phys., 35, 350, 1963.
\bibitem{Cannon}
Cannon, J. T. and Jordan, T. F., A No-Interaction Theorem in Classical Relativistic Hamiltonian Particle Dynamics, Journal of Math. Physics, 5, 299, 1964.
\bibitem{Leutwyler}
Leutwyler, M., A no-interaction theorem in classical relativistic Hamiltonian particle mechanics, Nuovo Cimento, 37, 556, 1965.
\bibitem{Magpantay1}
Magpantay, J. A., A Hamiltonian for Relativistic Interacting Many Particles, arXiv: 180602762 v1 [phys.class-ph].
\bibitem{Magpantay2} 
Magpantay, J. A., Does A Special Relativistic Liouville Equation Exist?, arxiv:2202.01951 v1 [cond-mat stat-mech].
\bibitem{Dirac}
Dirac, P. A. M., Lectures on Quantum Mechanics, Belfer Graduate School of Science, Yeshiva University, 1964.
\bibitem{Magpantay3}
Magpantay, J. A., A Special Relativistic Equation Exists, arxiv:2208.13387 v1 [cond-mat stat-mech]
\bibitem{Balescu}
Balescu, R., Relativistic Statistical Thermodynamics, Physica, 40, 309, 1968.
\bibitem{Tolman}
Tolman, R. C., Relativity, Thermodynamics and Cosmology, 1934, Oxford University Press.
\bibitem{Requardt}
Requardt, M., Thermodynamics meets Special Relativity - or what is real in Physics, arXiv:0801.2639v1 [gr-qc]
\bibitem{Landsberg}
Landsberg, P. T., Einstein and Statistical Thermodynamics I: Relativistic Thermodynamics, Eur. J. Phys. 2, 203, 1981.
\bibitem{Dunkel}
Dunkel, J., Hanggi, P. and Hilbert, S., Non-local observables and light-cone averaging in relativistic thermodynamics, Nature Physics DOI:10.1038, 2009.
\bibitem{Farias}
Farias, C., Pinto, V. and Moya, P., What is the temperature of a moving body?, Scientific Reports, Nature.com, 2017.
\bibitem{Wang}
Wang, C-Y, Thermodynamics Since Einstein, Advances in Natural Sciences, 6, 13, 2013.
\bibitem{deParga}
de Parga, G. A., Lopez-Carrera, B., Relativistic Statistical Mechanics vs Relativistic Thermodynamics, Entropy, 13, 1664, 2011.
\bibitem{Grandy}
Grandy, W. T., Entropy and the Time Evolution of Macroscopic Systems, Oxford University Press, 2008.
\bibitem{Quevedo}
Quevedo, H., Time is Entropy: A Geometric Proof, arXiv:2410.07639v1 [gr-qc]
\bibitem{Reif}
Reif, F., Fundamentals of Statistical and Thermal Physics, McGraw-Hill, Inc., 1965
\bibitem{Chacon-Acosta}
Chacon-Acosta, G. and Dagdug, L., On the Manifestly Covariant Juttner Distribution and Equipartition Theorem, arXiv: 09.10.1625v1 [cond-mat.stat-mech]
\end{thebibliography}
\end{document}